# ITO'S FORMULA FOR THE DISCRETE-TIME QUANTUM WALK IN TWO DIMENSIONS

**CLEMENT AMPADU**

31 Carrolton Road
Boston, Massachusetts, 02132
U.S.A.
e-mail: drampadu@hotmail.com

**Abstract**

Following [Konno, arXiv:1112.4335], it is natural to ask: What is the Ito's formula for the discrete time quantum walk on a graph different than $Z$, the set of integers? In this paper we answer the question for the discrete time quantum walk on $Z^2$, the square lattice.

## I. Introduction

Ito's formula which is related to the Ito's lemma is used in stochastic calculus to find the differential of a function of a particular type of stochastic process, and has a wide range of applications. According to the author in [1], the Ito's formula for the random walk has been investigated [2,3]. However, in the quantum walk case, the Ito's formulas are unknown. In [1] the author presents the Ito formula for the one-dimensional discrete-time quantum walk and gives some examples including Tanaka's formula by using the formula. Integrals for the quantum walk is also discussed.

In the present paper new results on the Ito's formula for the discrete-time quantum walk on the square lattice is given. This paper is organized as follows. In Section II we define the quantum walk on the square lattice, there the dynamics of the walk in the Fourier picture is recorded in Lemma 1. In Section III we present an Ito's formula for the discrete-time quantum walk on the square lattice as well as a Tanaka-type formula for the quantum walk. In [4] the author of the present paper computed some sojourn times of the Grover walk in two dimensions. The Tanaka-type formula may be useful in computing local time at the origin. Section IV is devoted to the conclusions, there two types of problems for further exploration is discussed. The first concerns integrals for the quantum walk whilst the second concerns another relation on the Ito's formula for the discrete-time quantum walk on the square lattice.

## II. Definitions

Recall that the discrete-time quantum walk is the quantum analogue of the classical random walk with an additional degree of freedom call chirality. In the two-dimensional setting on the square lattice, the chirality takes values left, right, downward, and upward, and means the direction of motion of the walker. At each time step, the particle moves according to its chirality state. For example, if the chirality state is upward, then the particle moves one step up. Let us define

$$|L\rangle = \begin{bmatrix} 1 \\ 0 \\ 0 \\ 0 \end{bmatrix}, |R\rangle = \begin{bmatrix} 0 \\ 1 \\ 0 \\ 0 \end{bmatrix}, |D\rangle = \begin{bmatrix} 0 \\ 0 \\ 1 \\ 0 \end{bmatrix}, \text{ and } |U\rangle = \begin{bmatrix} 0 \\ 0 \\ 0 \\ 1 \end{bmatrix}, \text{ where } L, R, D, U \text{ refer to the left, right, down,}$$

and up chirality states respectively. The time evolution of the quantum walk on $Z^2$ is determined by $U^{\otimes 2}$, where $U = \begin{bmatrix} a & b \\ c & d \end{bmatrix} \in U(2)$, with $a,b,c,d \in C$, where $C$ is the set of complex numbers. The unitarity of $U$ gives $|a|^2 + |b|^2 = |c|^2 + |d|^2 = 1$, $a\bar{c} + b\bar{d} = 0$, $c = -\Delta \bar{b}$, $d = \Delta \bar{a}$, where $\bar{z}$ denotes complex conjugation, and $\Delta = \det U = ad - bc$ with $|\Delta| = 1$. We should remark that $U^{\otimes 2}$ is a $4 \times 4$ matrix which is also unitary. In order to define the dynamics of the model we write

$$U^{\otimes 2} = P_{(-1,0)} + Q_{(1,0)} + R_{(0,-1)} + S_{(0,1)}, \text{ where } P_{(-1,0)} = \begin{bmatrix} a^2 & ab & ab & b^2 \\ 0 & 0 & 0 & 0 \\ 0 & 0 & 0 & 0 \\ 0 & 0 & 0 & 0 \end{bmatrix},$$

$$Q_{(1,0)} = \begin{bmatrix} 0 & 0 & 0 & 0 \\ ac & ad & bc & bd \\ 0 & 0 & 0 & 0 \\ 0 & 0 & 0 & 0 \end{bmatrix}, R_{(0,-1)} = \begin{bmatrix} 0 & 0 & 0 & 0 \\ 0 & 0 & 0 & 0 \\ ac & bc & ad & bd \\ 0 & 0 & 0 & 0 \end{bmatrix}, \text{ and } S_{(0,1)} = \begin{bmatrix} 0 & 0 & 0 & 0 \\ 0 & 0 & 0 & 0 \\ 0 & 0 & 0 & 0 \\ c^2 & cd & cd & d^2 \end{bmatrix}.$$

We should note that $P_{(-1,0)}, Q_{(1,0)}, R_{(0,-1)}, S_{(0,1)}$ represents the walker moves to the left, right, downward, and upward directions respectively at position $(x, y)$ at each time step. Let the set of initial quibit states at the origin for the quantum walk be given by

$$\Phi = \{\varphi = \alpha|L\rangle + \beta|R\rangle + \gamma|D\rangle + \lambda|U\rangle \in C^4 : |\alpha|^2 + |\beta|^2 + |\gamma|^2 + |\lambda|^2 = 1\}$$

$$= \{\varphi =^T [\alpha \ \beta \ \gamma \ \lambda] \in C^4 : |\alpha|^2 + |\beta|^2 + |\gamma|^2 + |\lambda|^2 = 1\}$$

Let $\Xi_n(l,r,d,u)$ denote the sum of all paths starting from the origin in the trajectory consisting of $l$ steps left, $r$ steps right, $d$ steps downwards, and $u$ steps upwards. For time $n = l + r + d + u$, and position $x = -l + r$, $y = -d + u$, we have

$\Xi_n(l,r,d,u) = \sum_{l_j,r_j,d_j,u_j} P^{l_1} Q^{r_1} R^{d_1} S^{u_1} \cdots P^{l_{n-1}} Q^{r_{n-1}} R^{d_{n-1}} S^{u_{n-1}} P^{l_n} Q^{r_n} R^{d_n} S^{u_n}$, where the summation is

taken over all integers $l_j, r_j, d_j, u_j \geq 0$ satisfying $\sum_{i=1}^n l_i = l$, $\sum_{i=1}^n r_i = r$, $\sum_{i=1}^n d_i = d$, $\sum_{i=1}^n u_i = u$, and

$l_j + r_j + d_j + u_j = 1$. We should note that the definition implies we can write

$\Xi_{n+1}(l,r,d,u) = P\Xi_n(l-1,r,d,u) + Q\Xi_n(l,r-1,d,u) + R\Xi_n(l,r,d-1,u) + S\Xi_n(l,r,d,u-1)$.

The probability that the quantum walker is in position $(x,y)$ at time $n$ starting from the origin with $\varphi \in \Phi$ is defined by $P(X_n = x, Y_n = y) = \|\Xi(l,r,d,u)\varphi\|^2$, where $n = l + r + d + u$, $x = -l + r$, and $y = -d + u$. The probability amplitude $\Psi_n(x,y)$ at position $(x,y)$ at time $n$ is given by

$\Psi_n(x,y) = \sum_{j \in \{L,R,D,U\}} \Psi_n^j(x,y)|j\rangle = \Xi_n(l,r,d,u)\varphi$. So, $P(X_n = x, Y_n = y) = \sum_{j \in \{L,R,D,U\}} |\Psi_n^j(x,y)|^2$. From

now on we consider the Fourier transform of $\Psi_n(x,y)$. By definition

$\Psi_n(x,y) = Q_{(1,0)}\Psi_n(x-1,y) + P_{(-1,0)}\Psi_n(x+1,y) + S_{(0,1)}\Psi_n(x,y-1) + R_{(0,-1)}\Psi_n(x,y+1)$. The Fourier transform of $\Psi_n^j(x,y)$ ($j = L, R, D, U$), that is, $\hat{\Psi}_n^j(\xi,\eta)$ is given by

$\hat{\Psi}_n^j(\xi,\eta) = \sum_{x,y \in \mathbb{Z}} e^{i\xi x + i\eta y} \Psi_n^j(x,y)$, from which it follows that

$\Psi_n^j(x,y) = \int_{-\pi}^{\pi}\int_{-\pi}^{\pi} \frac{d\xi}{2\pi} \frac{d\eta}{2\pi} e^{-i\xi x - i\eta y} \hat{\Psi}_n^j(\xi,\eta)$. Put $\hat{\Psi}_n(\xi,\eta) = \begin{bmatrix} \hat{\Psi}_n^L(\xi,\eta) \\ \hat{\Psi}_n^R(\xi,\eta) \\ \hat{\Psi}_n^D(\xi,\eta) \\ \hat{\Psi}_n^U(\xi,\eta) \end{bmatrix}$, then

$\hat{\Psi}_n(\xi,\eta) = \sum_{x,y \in \mathbb{Z}} e^{i\xi x + i\eta y} \Psi_n(x,y)$ and $\Psi_n(x,y) = \int_{-\pi}^{\pi}\int_{-\pi}^{\pi} \frac{d\xi}{2\pi} \frac{d\eta}{2\pi} e^{-i\xi x - i\eta y} \hat{\Psi}_n(\xi,\eta)$. From

$\Psi_n(x,y) = Q_{(1,0)}\Psi_n(x-1,y) + P_{(-1,0)}\Psi_n(x+1,y) + S_{(0,1)}\Psi_n(x,y-1) + R_{(0,-1)}\Psi_n(x,y+1)$, for $(\xi,\eta) \in [-\pi,\pi) \times [-\pi,\pi)$, we obtain the following

**Lemma 1:** For any $n = 0,1,2,\ldots$

$\hat{\Psi}_{n+1}(\xi,\eta) = U(\xi,\eta)\hat{\Psi}_n(\xi,\eta)$. Here $U(\xi,\eta)$ is given by

$U(\xi,\eta)$
$= e^{-i\xi} P_{(-1,0)} + e^{i\xi} Q_{(1,0)} + e^{-i\eta} R_{(0,-1)} + e^{i\eta} S_{(0,1)}$
$= Diag(e^{-i\xi}, e^{i\xi}, e^{-i\eta}, e^{i\eta})U$

Note that $\hat{\Psi}_0(\xi,\eta) = \begin{bmatrix} \alpha \\ \beta \\ \gamma \\ \lambda \end{bmatrix} \in C^4$, where $|\alpha|^2 + |\beta|^2 + |\gamma|^2 + |\lambda|^2 = 1$. In terms of $\hat{\Psi}_0(\xi,\eta)$ it follows by induction on $n$ from Lemma 1 that $\hat{\Psi}_n(\xi,\eta) = U(\xi,\eta)^n \hat{\Psi}_0(\xi,\eta)$. Note that

$$P(X_n = x, Y_n = y) = \|\Psi_n(x,y)\|^2 = \left( \int_{-\pi}^{\pi}\int_{-\pi}^{\pi} \frac{d\xi}{2\pi} \frac{d\eta}{2\pi} e^{-i\xi x - i\eta y} U(\xi,\eta)^n \hat{\Psi}_0(\xi,\eta) \right)^* \int_{-\pi}^{\pi}\int_{-\pi}^{\pi} \frac{d\xi}{2\pi} \frac{d\eta}{2\pi} e^{-i\xi x - i\eta y} U(\xi,\eta)^n \hat{\Psi}_0(\xi,\eta)$$

, where * means the adjoint operator. Write

$$\left( \int_{-\pi}^{\pi}\int_{-\pi}^{\pi} \frac{d\xi}{2\pi} \frac{d\eta}{2\pi} e^{-i\xi x - i\eta y} U(\xi,\eta)^n \hat{\Psi}_0(\xi,\eta) \right)^* = \int_{-\pi}^{\pi}\int_{-\pi}^{\pi} \frac{d\xi'}{2\pi} \frac{d\eta'}{2\pi} e^{i\xi' x + i\eta' y} U(\xi',\eta')^n \hat{\Psi}_0(\xi',\eta')$$, then,

$$P(X_n = x, Y_n = y) = \|\Psi_n(x,y)\|^2 = \int_{-\pi}^{\pi}\int_{-\pi}^{\pi} \frac{d\xi'}{2\pi} \frac{d\eta'}{2\pi} e^{i\xi' x + i\eta' y} U(\xi',\eta')^n \hat{\Psi}_0(\xi',\eta') \int_{-\pi}^{\pi}\int_{-\pi}^{\pi} \frac{d\xi}{2\pi} \frac{d\eta}{2\pi} e^{-i\xi x - i\eta y} U(\xi,\eta)^n \hat{\Psi}_0(\xi,\eta)$$

Since $U(\xi,\eta)$ is a power of $n$, in order to calculate $P(X_n = x, Y_n = y)$ it is usual practice to diagonalize the matrix $U(\xi,\eta)$. We should note that $U(\xi',\eta')$ is also a power of $n$ by the adjoint operation.

### III. Ito-Type Formulas for the Discrete-Time Quantum Walk on the Square Lattice

Let $B_n = \{-n, -(n-1), \ldots, n-1, n\}$, $B_{n'} = \{-n', -(n'-1), \ldots, n'-1, n'\}$, $\Omega_n = B_n^{n+1}$, $\Omega_{n'} = B_{n'}^{n'+1}$,

$w_n = (w_n(0) = 0, w_n(1), w_n(2), \ldots, w_n(n)) \in \Omega_n$,

$w_{n'} = (w_{n'}(0) = 0, w_{n'}(1), w_{n'}(2), \ldots, w_{n'}(n')) \in \Omega_{n'}$,

$v_n = (w_n(1), w_n(2) - w_n(1), \ldots, w_n(n) - w_n(n-1))$

$v_{n'} = (w_{n'}(1), w_{n'}(2) - w_{n'}(1), \ldots, w_{n'}(n') - w_{n'}(n'-1))$,

$u_n = (I_{\{1\}}(v_n(1)), I_{\{1\}}(v_n(2)), \ldots, I_{\{1\}}(v_n(n)))$, $u_{n'} = (I_{\{1\}}(v_{n'}(1)), I_{\{1\}}(v_{n'}(2)), \ldots, I_{\{1\}}(v_{n'}(n')))$, where $I_A(x)$ indicates the indicator function for a set $A$. Note that $w_n(m+1) - w_n(m), w_{n'}(m'+1) - w_{n'}(m') \in \{-1,1\}$. From now on we consider the quantum walk on $\Omega_{n,n'} = B_n^{n+1} \times B_{n'}^{n'+1}$. To do so let $w_{n,n'} = w_n \times w_{n'} \in \Omega_{n,n'}$, $v_{n,n'} = v_n \times v_{n'}$, $u_{n,n'} = u_n \times u_{n'}$, and noting that $w_{n,n'}(m+1,m') - w_{n,n'}(m,m') \in \{-1,1\} \times \{-1,1\}$ as well as $w_{n,n'}(m,m'+1) - w_{n,n'}(m,m') \in \{-1,1\} \times \{-1,1\}$, a direct computation gives the following

**Proposition 2:** Let $f : Z^2 \to C$. For any $m \in \{0,1,\ldots,n-1\}$ and $m' \in \{0,1,\ldots,n'-1\}$ we have

a)

$$f(w_{n,n'}(m+1,m')) - f(w_{n,n'}(m,m')) = \frac{1}{2}\{f(w_{n,n'}(m,m')+1) - f(w_{n,n'}(m,m')-1)\}(w_{n,n'}(m+1,m') - w_{n,n'}(m,m'))$$

$$+ \frac{1}{2}\{f(w_{n,n'}(m,m')+1) - 2f(w_{n,n'}(m,m')) + f(w_{n,n'}(m,m')-1)\}$$

$$f(w_{n,n'}(m,m'+1)) - f(w_{n,n'}(m,m')) = \frac{1}{2}\{f(w_{n,n'}(m,m')+1) - f(w_{n,n'}(m,m')-1)\}(w_{n,n'}(m,m'+1) - w_{n,n'}(m,m'))$$

$$+ \frac{1}{2}\{f(w_{n,n'}(m,m')+1) - 2f(w_{n,n'}(m,m')) + f(w_{n,n'}(m,m')-1)\}$$

b)

$$f(w_{n,n'}(n,n')) - f(w_{n,n'}(0,n')) = \frac{1}{2}\sum_{m=0}^{n-1}\{f(w_{n,n'}(m,m')+1) - f(w_{n,n'}(m,m')-1)\}(w_{n,n'}(m+1,m') - w_{n,n'}(m,m'))$$

$$+ \frac{1}{2}\sum_{m=0}^{n-1}\{f(w_{n,n'}(m,m')+1) - 2f(w_{n,n'}(m,m')) + f(w_{n,n'}(m,m')-1)\}$$

$$f(w_{n,n'}(n,n')) - f(w_{n,n'}(n,0)) = \frac{1}{2}\sum_{m'=0}^{n-1}\{f(w_{n,n'}(m,m')+1) - f(w_{n,n'}(m,m')-1)\}(w_{n,n'}(m,m'+1) - w_{n,n'}(m,m')) +$$

$$\frac{1}{2}\sum_{m'=0}^{n-1}\{f(w_{n,n'}(m,m')+1) - 2f(w_{n,n'}(m,m')) + f(w_{n,n'}(m,m')-1)\}$$

**Proof: (b)** can easily been seen by summing the expressions in **(a)** from 0 to $n-1$, with respect to the indices $m$ and $m'$ to get the first and second expressions respectively. In particular, it is easy to see that the right-hand side of the expressions in **(b)** have been left in terms of the sigma notation, where the sum over the appropriate indices run from 0 to $n-1$. After summing the expressions in **(a)** from 0 to $n-1$, with respect to the indices $m$ and $m'$, the left hand side becomes a telescoping sum, that is, the sum collapses to just two terms as can been seen in the left hand side of the expressions in **(b)** . To see the expressions in **(a)**, for example the first one, let $w_{n,n'}(m+1,m') - w_{n,n'}(m,m') = 1$, then after some algebra we get

$RHS = f(w_{n,n'}(m,m')+1) - f(w_{n,n'}(m,m'))$, however we have let

$w_{n,n'}(m+1,m') - w_{n,n'}(m,m') = 1$, so $w_{n,n'}(m+1,m') = w_{n,n'}(m,m')+1$, thus we have

$$RHS = f(w_{n,n'}(m,m')+1) - f(w_{n,n'}(m,m'))$$
$$= f(w_{n,n'}(m+1,m')) - f(w_{n,n'}(m,m'))$$

, and the proof is finished.

We should remark that the expressions in part **(A)** of Proposition 2 is the Ito formulas for the discrete-time quantum walk on the square lattice.

Put $k = u_n(n)2^{n-1} + u_n(n-1)2^{n-2} + \ldots + u_n(2)2^1 + u_n(1)2^0$, and
$k' = u_{n'}(n')2^{n'-1} + u_{n'}(n'-1)2^{n'-2} + \ldots + u_{n'}(2)2^1 + u_{n'}(1)2^0$, and let $P_{w_{n,n'}^{(k,k')}} = P_{w_n^{(k)}} \otimes P_{w_{n'}^{(k')}}$, where

$P_{w_n^{(k)}} = P_{v_n^{(k)}(n)} \cdots P_{v_n^{(k)}(2)} P_{v_n^{(k)}(1)}$, and $P_{w_{n'}^{(k')}}$ is defined in a similar way, then from Proposition 2 we immediately obtain the following

**Theorem 3:** Let $f : Z^2 \to C$. For any $m \in \{0,1,\ldots,n-1\}$ and $m' \in \{0,1,\ldots,n'-1\}$ we have,

**(a)**

$$\sum_{k=0}^{2^n-1}\sum_{k'=0}^{2^{n'}-1} \{f(w_{n,n'}^{(k,k')}(m+1,m')) - f(w_{n,n'}^{(k,k')}(m,m'))\}P_{w_{n,n'}^{(k,k')}} = \frac{1}{2}\sum_{k=0}^{2^n-1}\sum_{k'=0}^{2^{n'}-1}\{f(w_{n,n'}^{(k,k')}(m,m')+1) - f(w_{n,n'}^{(k,k')}(m,m')-1)\}(w_{n,n'}^{(k,k')}(m+1,m') - w_{n,n'}^{(k,k')}(m,m'))P_{w_{n,n'}^{(k,k')}} +$$
$$\frac{1}{2}\sum_{k=0}^{2^n-1}\sum_{k'=0}^{2^{n'}-1}\{f(w_{n,n'}^{(k,k')}(m,m')+1) - 2f(w_{n,n'}^{(k,k')}(m,m')) + f(w_{n,n'}^{(k,k')}(m,m')-1)\}P_{w_{n,n'}^{(k,k')}}$$

$$\sum_{k=0}^{2^n-1}\sum_{k'=0}^{2^{n'}-1} \{f(w_{n,n'}^{(k,k')}(m,m'+1)) - f(w_{n,n'}^{(k,k')}(m,m'))\}P_{w_{n,n'}^{(k,k')}} = \frac{1}{2}\sum_{k=0}^{2^n-1}\sum_{k'=0}^{2^{n'}-1}\{f(w_{n,n'}^{(k,k')}(m,m')+1) - f(w_{n,n'}^{(k,k')}(m,m')-1)\}(w_{n,n'}^{(k,k')}(m,m'+1) - w_{n,n'}^{(k,k')}(m,m'))P_{w_{n,n'}^{(k,k')}} +$$
$$\frac{1}{2}\sum_{k=0}^{2^n-1}\sum_{k'=0}^{2^{n'}-1}\{f(w_{n,n'}^{(k,k')}(m,m')+1) - 2f(w_{n,n'}^{(k,k')}(m,m')) + f(w_{n,n'}^{(k,k')}(m,m')-1)\}P_{w_{n,n'}^{(k,k')}}$$

**(b)**

$$\sum_{k=0}^{2^n-1}\sum_{k'=0}^{2^{n'}-1} \{f(w_{n,n'}^{(k,k')}(n',n)) - f(w_{n,n'}^{(k,k')}(0,n'))\}P_{w_{n,n'}^{(k,k')}} = \frac{1}{2}\sum_{k=0}^{2^n-1}\sum_{k'=0}^{2^{n'}-1}\sum_{m=0}^{n-1}\sum_{m'=0}^{n'-1}\{f(w_{n,n'}^{(k,k')}(m,m')+1) - f(w_{n,n'}^{(k,k')}(m,m')-1)\}(w_{n,n'}^{(k,k')}(m+1,m') - w_{n,n'}^{(k,k')}(m,m'))P_{w_{n,n'}^{(k,k')}} +$$
$$\frac{1}{2}\sum_{k=0}^{2^n-1}\sum_{k'=0}^{2^{n'}-1}\sum_{m=0}^{n-1}\sum_{m'=0}^{n'-1}\{f(w_{n,n'}^{(k,k')}(m,m')+1) - 2f(w_{n,n'}^{(k,k')}(m,m')) + f(w_{n,n'}^{(k,k')}(m,m')-1)\}P_{w_{n,n'}^{(k,k')}}$$

$$\sum_{k=0}^{2^n-1}\sum_{k'=0}^{2^{n'}-1} \{f(w_{n,n'}^{(k,k')}(n',n)) - f(w_{n,n'}^{(k,k')}(n,0))\}P_{w_{n,n'}^{(k,k')}} = \frac{1}{2}\sum_{k=0}^{2^n-1}\sum_{k'=0}^{2^{n'}-1}\sum_{m=0}^{n-1}\sum_{m'=0}^{n'-1}\{f(w_{n,n'}^{(k,k')}(m,m')+1) - f(w_{n,n'}^{(k,k')}(m,m')-1)\}(w_{n,n'}^{(k,k')}(m+1,m') - w_{n,n'}^{(k,k')}(m,m'))P_{w_{n,n'}^{(k,k')}} +$$
$$\frac{1}{2}\sum_{k=0}^{2^n-1}\sum_{k'=0}^{2^{n'}-1}\sum_{m=0}^{n-1}\sum_{m'=0}^{n'-1}\{f(w_{n,n'}^{(k,k')}(m,m')+1) - 2f(w_{n,n'}^{(k,k')}(m,m')) + f(w_{n,n'}^{(k,k')}(m,m')-1)\}P_{w_{n,n'}^{(k,k')}}$$

We should remark that when we consider $P_{(-1,0)} \to p \in [0,1]$, $Q_{(1,0)} \to q \in [0,1]$, $R_{(0,-1)} \to r \in [0,1]$, and $S_{(0,1)} \to s \in [0,1]$, with $p+q+r+s=1$, where $p,q,r,s$ corresponds to the probabilities of the walker moving left, right, down, and up in a random walk, then Theorem 3 is the corresponding result for the random walk on the square lattice which is *not* symmetric. If $p = q = r = s = \dfrac{1}{4}$, then the results corresponds to the simple symmetric random walk on the square lattice.

Next we present a Tanaka-type formula for the discrete-time quantum walk on the square lattice.

From first expression in part **(A)** of Theorem 3, we put $f\left(w_{n,n'}^{(k,k')}(m+1,m')\right) = \left|w_{n,n'}^{(k,k')}(m+1,m')\right|$,

$f(w_{n,n'}^{(k,k')}(m,m')) = 0$ for any $k$, $\dfrac{f\left(w_{n,n'}^{(k,k')}(m,m')+1\right) - f\left(w_{n,n'}^{(k,k')}(m,m')-1\right)}{2} = \mathrm{sgn}\left(w_{n,n'}^{(k,k')}(m,m')\right)$

where $\mathrm{sgn}(\cdot)$ denotes the sign function,

$$\dfrac{f\left(w_{n,n'}^{(k,k')}(m,m')+1\right) - 2f(w_{n,n'}^{(k,k')}(m,m')) + f\left(w_{n,n'}^{(k,k')}(m,m')-1\right)}{2} = I_{\{0\}}\left(w_{n,n'}^{(k,k')}(m,m')\right).$$

Making a similar substitution in the second expression in part **(A)** of Theorem 3, we get the Tanaka-type formula for the discrete-time quantum walk on the square lattice as follows.

**Corollary 4:**

**(a)**

$$\sum_{k=0}^{2^n-1}\sum_{k'=0}^{2^{n'}-1}\left|w_{n,n'}^{(k,k')}(m+1,m')\right|P_{w_{n,n'}^{(k,k')}} = \sum_{k=0}^{2^n-1}\sum_{k'=0}^{2^{n'}-1}\mathrm{sgn}\left(w_{n,n'}^{(k,k')}(m,m')\right)\left(w_{n,n'}^{(k,k')}(m+1,m') - w_{n,n'}^{(k,k')}(m,m')\right)P_{w_{n,n'}^{(k,k')}} + \sum_{k=0}^{2^n-1}\sum_{k'=0}^{2^{n'}-1}I_{\{0\}}\left(w_{n,n'}^{(k,k')}(m,m')\right)P_{w_{n,n'}^{(k,k')}}$$

**(b)**

$$\sum_{k=0}^{2^n-1}\sum_{k'=0}^{2^{n'}-1}\left|w_{n,n'}^{(k,k')}(m,m'+1)\right|P_{w_{n,n'}^{(k,k')}} = \sum_{k=0}^{2^n-1}\sum_{k'=0}^{2^{n'}-1}\mathrm{sgn}\left(w_{n,n'}^{(k,k')}(m,m')\right)\left(w_{n,n'}^{(k,k')}(m+1,m') - w_{n,n'}^{(k,k')}(m,m')\right)P_{w_{n,n'}^{(k,k')}} + \sum_{k=0}^{2^n-1}\sum_{k'=0}^{2^{n'}-1}I_{\{0\}}\left(w_{n,n'}^{(k,k')}(m,m')\right)P_{w_{n,n'}^{(k,k')}}$$

Recall to compute $P(X_n = x, Y_n = y)$ it is necessary to diagonalize $U(\xi,\eta)$ since it is a power of $n$ in the expression for $P(X_n = x, Y_n = y)$. Next we give a formula that slightly expands $U(\xi,\eta)^n$ making use of Theorem 3.

Let $f\left(w_{n,n'}^{(k,k')}(n',n)\right) = e^{i\xi w_{n,n'}^{(k,k')}(n',n) + i\eta w_{n,n'}^{(k,k')}(n',n)}$ and $f(w_{n,n'}^{(k,k')}(0,n')) = 1$ for any $k, k',$ and $n'$, then the LHS of the first expression in part **(b)** of Theorem 3 becomes

$$\sum_{k=0}^{2^n-1}\sum_{k'=0}^{2^{n'}-1}\left\{e^{i\xi w_{n,n'}^{(k,k')}(n',n)+i\eta w_{n,n'}^{(k,k')}(n',n)} - 1\right\}P_{w_{n,n'}^{(k,k')}} = U(\xi,\eta)^n - U^n.$$ As for the RHS of the first expression in part **(b)** of Theorem 3, we can write the first term as

$$\frac{1}{2}\sum_{k=0}^{2^n-1}\sum_{k'=0}^{2^{n'}-1}\sum_{m=0}^{n-1}\sum_{m'=0}^{n'-1}\{f(w_{n,n'}^{(k,k')}(m,m')+1)-f(w_{n,n'}^{(k,k')}(m,m')-1)\}(w_{n,n'}^{(k,k')}(m+1,m')-w_{n,n'}^{(k,k')}(m,m'))P_{w_{n,n'}^{(k,k')}}$$

$$=\frac{1}{2}\sum_{k=0}^{2^n-1}\sum_{k'=0}^{2^{n'}-1}\sum_{m=0}^{n-1}\sum_{m'=0}^{n'-1}\{e^{i\xi w_{n,n'}^{(k,k')}(m',m')+i\eta w_{n,n'}^{(k,k')}(m',m)}\left(e^{i(\xi+\eta)}-e^{-i(\xi+\eta)}\right)\}(w_{n,n'}^{(k,k')}(m+1,m')-w_{n,n'}^{(k,k')}(m,m'))P_{w_{n,n'}^{(k,k')}}$$

$$=i\sin(\xi+\eta)\sum_{k=0}^{2^n-1}\sum_{k'=0}^{2^{n'}-1}\sum_{m=0}^{n-1}\sum_{m'=0}^{n'-1}\{e^{i\xi w_{n,n'}^{(k,k')}(m',m')+i\eta w_{n,n'}^{(k,k')}(m',m)}\}(w_{n,n'}^{(k,k')}(m+1,m')-w_{n,n'}^{(k,k')}(m,m'))P_{w_{n,n'}^{(k,k')}}$$

Similarly, we can show that the second term on the RHS of the first expression in part **(b)** of Theorem 3, can be written as

$$\frac{1}{2}\sum_{k=0}^{2^n-1}\sum_{k'=0}^{2^{n'}-1}\sum_{m=0}^{n-1}\sum_{m'=0}^{n'-1}\{f(w_{n,n'}^{(k,k')}(m,m')+1)-2f(w_{n,n'}^{(k,k')}(m,m'))+f(w_{n,n'}^{(k,k')}(m,m')-1)\}P_{w_{n,n'}^{(k,k')}}$$

$$=(\cos(\xi+\eta)-1)\sum_{k=0}^{2^n-1}\sum_{k'=0}^{2^{n'}-1}\sum_{m=0}^{n-1}\sum_{m'=0}^{n'-1}e^{i\xi w_{n,n'}^{(k,k')}(m',m')+i\eta w_{n,n'}^{(k,k')}(m',m)}P_{w_{n,n'}^{(k,k')}}$$

So the RHS of the first expression in part **(b)** of Theorem 3 becomes

$$i\sin(\xi+\eta)\sum_{k=0}^{2^n-1}\sum_{k'=0}^{2^{n'}-1}\sum_{m=0}^{n-1}\sum_{m'=0}^{n'-1}\{e^{i\xi w_{n,n'}^{(k,k')}(m',m')+i\eta w_{n,n'}^{(k,k')}(m',m)}\}(w_{n,n'}^{(k,k')}(m+1,m')-w_{n,n'}^{(k,k')}(m,m'))P_{w_{n,n'}^{(k,k')}}$$

$$+(\cos(\xi+\eta)-1)\sum_{k=0}^{2^n-1}\sum_{k'=0}^{2^{n'}-1}\sum_{m=0}^{n-1}\sum_{m'=0}^{n'-1}e^{i\xi w_{n,n'}^{(k,k')}(m',m')+i\eta w_{n,n'}^{(k,k')}(m',m)}P_{w_{n,n'}^{(k,k')}}$$

Now equating the expression immediately above to $U(\xi,\eta)^n - U^n$, we get

$$U(\xi,\eta)^n = U^n + i\sin(\xi+\eta)\sum_{k=0}^{2^n-1}\sum_{k'=0}^{2^{n'}-1}\sum_{m=0}^{n-1}\sum_{m'=0}^{n'-1}\{e^{i\xi w_{n,n'}^{(k,k')}(m',m')+i\eta w_{n,n'}^{(k,k')}(m',m)}\}(w_{n,n'}^{(k,k')}(m+1,m')-w_{n,n'}^{(k,k')}(m,m'))P_{w_{n,n'}^{(k,k')}}$$

$$+(\cos(\xi+\eta)-1)\sum_{k=0}^{2^n-1}\sum_{k'=0}^{2^{n'}-1}\sum_{m=0}^{n-1}\sum_{m'=0}^{n'-1}e^{i\xi w_{n,n'}^{(k,k')}(m',m')+i\eta w_{n,n'}^{(k,k')}(m',m)}P_{w_{n,n'}^{(k,k')}}$$

Repeating the argument immediately above to the LHS and RHS of the second expression in part **(b)** of Theorem 3, we also get

$$U(\xi,\eta)^n = U^n + i\sin(\xi+\eta)\sum_{k=0}^{2^n-1}\sum_{k'=0}^{2^{n'}-1}\sum_{m=0}^{n-1}\sum_{m'=0}^{n'-1}\left\{e^{i\xi w_{n,n'}^{(k,k')}(m',m')+i\eta w_{n,n'}^{(k,k')}(m',m)}\right\}\left(w_{n,n'}^{(k,k')}(m,m'+1)-w_{n,n'}^{(k,k')}(m,m')\right)P_{w_{n,n'}^{(k,k')}}$$

$$+(\cos(\xi+\eta)-1)\sum_{k=0}^{2^n-1}\sum_{k'=0}^{2^{n'}-1}\sum_{m=0}^{n-1}\sum_{m'=0}^{n'-1}e^{i\xi w_{n,n'}^{(k,k')}(m',m')+i\eta w_{n,n'}^{(k,k')}(m',m)}P_{w_{n,n'}^{(k,k')}}$$

So from the expressions immediately above we have the following

**Corollary 5:**

$$U(\xi,\eta)^n = U^n + i\sin(\xi+\eta)\sum_{k=0}^{2^n-1}\sum_{k'=0}^{2^{n'}-1}\sum_{m=0}^{n-1}\sum_{m'=0}^{n'-1}\left\{e^{i\xi w_{n,n'}^{(k,k')}(m',m')+i\eta w_{n,n'}^{(k,k')}(m',m)}\right\}\left(w_{n,n'}^{(k,k')}(m+1,m')-w_{n,n'}^{(k,k')}(m,m')\right)P_{w_{n,n'}^{(k,k')}}$$

$$+(\cos(\xi+\eta)-1)\sum_{k=0}^{2^n-1}\sum_{k'=0}^{2^{n'}-1}\sum_{m=0}^{n-1}\sum_{m'=0}^{n'-1}e^{i\xi w_{n,n'}^{(k,k')}(m',m')+i\eta w_{n,n'}^{(k,k')}(m',m)}P_{w_{n,n'}^{(k,k')}}$$

$$U(\xi,\eta)^n = U^n + i\sin(\xi+\eta)\sum_{k=0}^{2^n-1}\sum_{k'=0}^{2^{n'}-1}\sum_{m=0}^{n-1}\sum_{m'=0}^{n'-1}\left\{e^{i\xi w_{n,n'}^{(k,k')}(m',m')+i\eta w_{n,n'}^{(k,k')}(m',m)}\right\}\left(w_{n,n'}^{(k,k')}(m,m'+1)-w_{n,n'}^{(k,k')}(m,m')\right)P_{w_{n,n'}^{(k,k')}}$$

$$+(\cos(\xi+\eta)-1)\sum_{k=0}^{2^n-1}\sum_{k'=0}^{2^{n'}-1}\sum_{m=0}^{n-1}\sum_{m'=0}^{n'-1}e^{i\xi w_{n,n'}^{(k,k')}(m',m')+i\eta w_{n,n'}^{(k,k')}(m',m)}P_{w_{n,n'}^{(k,k')}}$$

**IV. Concluding Remarks**

In this paper we have shown the Ito's formula for the discrete-time quantum walk on the square lattice and as a consequence obtained a Tanaka-type formula for the quantum walk. The relation to the simple random walk on the square lattice has also been clarified in the biased and unbiased case. Following the author in [1] it is an interesting problem to clarify the relation between $\sigma_{n,n'}(f) = \sum_{k=0}^{2^n-1}\sum_{k'=0}^{2^{n'}-1} f\left(w_{n,n'}^{(k,k')}\right) P_{w_{n,n'}^{(k,k')}}$ and $\iint f d\mu_{n,n'}$. If $f$ is a function of two variables, $f(x,y)$. It is well known that $\Delta f = f_x \Delta x + f_y \Delta y$, exploiting this relation another Ito formula for the discrete-time quantum walk on the square lattice is the following. We record as a conjecture, therefore one of the future interesting problems is to verify or refute the following

**Conjecture 6:** Let $f : Z^2 \to C$. For any $m \in \{0,1,\ldots,n-1\}$ and $m' \in \{0,1,\ldots,n'-1\}$, we have

$$f(w_{n,n'}(m+1,m'+1)) - f(w_{n,n'}(m,m')) =$$

$$\frac{1}{2}\{f(w_{n,n'}(m,m'+1)+1) - f(w_{n,n'}(m,m'-1)-1)\}(w_{n,n'}(m,m'+1) - w_{n,n'}(m,m')) +$$

$$\frac{1}{2}\{f(w_{n,n'}(m+1,m')+1) - f(w_{n,n'}(m-1,m')-1)\}(w_{n,n'}(m+1,m') - w_{n,n'}(m,m')) +$$

$$\frac{1}{2}\{f(w_{n,n'}(m,m'+1)+1) + f(w_{n,n'}(m,m'-1)-1) - 4f(w_{n,n'}(m,m')) + f(w_{n,n'}(m+1,m')+1) - f(w_{n,n'}(m-1,m')-1)\}$$